\documentclass[letterpaper, 10 pt, conference]{IEEEtran}
\IEEEoverridecommandlockouts
\usepackage{cite}
\usepackage{amsmath,amssymb,amsfonts}
\usepackage{algorithmic}
\usepackage{graphicx}
\usepackage{textcomp}
\usepackage{float}
\usepackage{graphicx}
\usepackage{mathtools}
\usepackage{hyperref} 
\usepackage{xcolor}
\usepackage{balance}
\usepackage{amsmath,amssymb,nicefrac,braket,graphicx,subcaption,xfrac,dirtytalk}
\allowdisplaybreaks
\def\BibTeX{{\rm B\kern-.05em{\sc i\kern-.025em b}\kern-.08em
		T\kern-.1667em\lower.7ex\hbox{E}\kern-.125emX}}
\begin{document}
	
	\newcommand{\s}{\boldsymbol{s}}
	\newcommand{\sk}{\boldsymbol{s}^{(k)}}
	\newcommand{\skp}{\boldsymbol{s}^{(k+1)}}
	\newcommand{\es}{E(\boldsymbol{s})}
	\newcommand{\esk}{E(\boldsymbol{s}^{(k)})}
	\newcommand{\eskp}{E(\boldsymbol{s}^{(k+1)})}
	\newcommand{\des}{E(\boldsymbol{s})-E^*}
	\newcommand{\desk}{E(\boldsymbol{s}^{(k)})-E^*}
	\newcommand{\deskp}{E(\boldsymbol{s}^{(k+1)})-E^*}
	\newcommand{\fk}{\boldsymbol{f}^{(k)}}
	\newcommand{\fkp}{\boldsymbol{f}^{(k+1)}}
	\newcommand{\nk}{\boldsymbol{\zeta}^{(k)}}
	\newcommand{\J}{\boldsymbol{J}}
	\newcommand{\h}{\boldsymbol{h}}
	\newcommand{\Q}{\boldsymbol{Q}}
	\newcommand{\ges}{\nabla E(\boldsymbol{s})}
	\newcommand{\gesk}{\nabla E(\boldsymbol{s}^{(k)})}
	\newcommand{\geskp}{\nabla E(\boldsymbol{s}^{(k+1)})}
	\newcommand{\F}{\mathcal{F}}
	\newcommand{\edesk}{\mathbb{E}[\desk]}
	\newcommand{\edeskp}{\mathbb{E}[\deskp]}
	\newcommand{\cedesk}{\mathbb{E}[\desk \vert \mathcal{F}_k]}
	\newcommand{\cedeskp}{\mathbb{E}[\deskp \vert \mathcal{F}_k]}
	\newcommand{\ex}{\mathbb{E}}
	
	\title{Convergence Analysis of Opto-Electronic Oscillator based Coherent Ising Machines}
	
	\author{
		\IEEEauthorblockN{Sayantan Pramanik$^{1,2}$, Sourav Chatterjee$^{1}$, Harshkumar Oza$^{3*}$\thanks{$^*$Work done during internship at TCS}}
		\IEEEauthorblockA{$^1$Corporate Incubation, TCS Research \& Innvovation, TATA Consultancy Services \\ $^2$Robert Bosch Centre for Cyber-Physical Systems, Indian Institute of Science \\ $^3$Department of Instrumentation and Applied Physics, Indian Institute of Science \\
			sayantan.pramanik@tcs.com, sourav.chat@tcs.com, ozah@iisc.ac.in}
	}
	
	\maketitle
	\begin{abstract}
		Ising machines are purported to be better at solving large-scale combinatorial optimisation problems better than conventional von Neumann computers. However, these Ising machines are widely believed to be heuristics, whose promise is observed empirically rather than obtained theoretically. We bridge this gap by considering an opto-electronic oscillator based coherent Ising machine, and providing the first analytical proof that under reasonable assumptions, the OEO-CIM is not a heuristic approach. We find and prove bounds on its performance in terms of the expected difference between the objective value at the final iteration and the optimal one, and on the number of iterations required by it. In the process, we emphasise on some of its limitations such as the inability to handle asymmetric coupling between spins, and the absence of external magnetic field applied on them (both of which are necessary in many optimisation problems), along with some issues in its convergence. We overcome these limitations by proposing suitable adjustments and prove that the improved architecture is guaranteed to converge to the optimum of the relaxed objective function.
	\end{abstract}
	
	\section{Introduction}\label{sec:intro}
	
	Recent developments in photonics-based analog computing have triggered a global exigency towards realizing new computational paradigms\,\cite{abel2019silicon, yamamoto2020coherent, mourgias2023analog} that can potentially outperform conventional digital computers in executing challenging computational tasks, such as solving combinatorial optimisation problems that are known to be NP-hard\,\cite{honjo2021100, mcmahon2016fully, mourgias2023analog}. Many NP-hard combinatorial optimisation problems can be efficiently mapped to a ground-state-search problem of the Ising model\,\cite{ising_np} -- which is basically a mathematical abstraction of magnetic systems describing the behavior of competitively interacting spins or angular momenta of fundamental particles\,\cite{lenz1920beitrag, ising1925beitrag}. Such an Ising model of coupled artificial spins can be physically realized using various systems ranging from Josephson junction\,\cite{barends2016digitized}, trapped ions\,\cite{kim2010quantum}, to optical states\,\cite{wang2013coherent}. Among these possible realizations, the optics-based approach involving coherent Ising machines (CIMs) is advantageous because of operability at room temperature, the requirement of cost-effective off-the-shelf components, miniaturisation possibilities over integrated photonic circuits, and the scope to implement dense and flexible coupling topologies\,\cite{yamamoto2017coherent, bohm2019poor}. Furthermore, being intrinsically gain-dissipative systems, CIMs are naturally prone towards avoiding local energy minima and thus achieving optimal solutions\,\cite{bohm2019poor,leleu2017combinatorial}. Given a very short-lived evolution to the ground state or optimal solution, the Ising machines promise significant speed up over conventional (digital) computers in solving hard optimisation problems\,\cite{inagaki2016coherent, bohm2019poor}, efficient solution of which is central to several industry-verticals\,\cite{bohm2019poor, mourgias2023analog}. 
	
	CIMs belong to the non-von Neumann computing architecture. They compute very differently from conventional (digital) computers, digital annealers, quantum annealers, or even gate-based quantum computers. In that light, they exhibit a few distinct advantages over many other computing approaches. This includes higher speed and efficiency over classical computing, for instance: a few hundred THz bandwidth in optical compared to that of a few GHz in classical computing; along with inherent parallel operations and in-memory computation at lower energy costs. Next, as opposed to the alleged effectiveness of the quantum approximate optimisation algorithm (QAOA)\,\cite{qaoa} in efficiently solving NP-hard optimisation problems against commercial solvers executed on digital computers, the benchmarking results obtained with CIM are more promising and convincing\,\cite{yamamoto2018optical}. Furthermore, unlike the gate and annealing based model of quantum computation, the operation of CIMs is not limited by factors such as scalability, high noise, and cryogenic cooling\,\cite{yamamoto2018optical}. Lastly, in absence of noise the anneal time can be significantly slower for quantum annealing based approaches which is not the case for CIMs -- being optically driven it is not subjected to thermal noise. Finally, digital annealers can be potentially outperformed by CIMs, since unlike CIMs they are practically limited by digital CMOS hardware-constraints. 

	Quite a few methods of experimentally realizing CIMs have been reported in literature. They all commonly use optical setups along with programmable fast feedback electronics to implement Ising spin networks with bistable coherent optical states\,\cite{bohm2019poor}. This enables them to naturally approach the optimal states of the programmed Hamiltonian. These methods of realizing Ising spin systems with a network of coupled optical states can be broadly classified into three types: (a) CIM based on mutually coupled lasers\,\cite{utsunomiya2015binary} (b) degenerate optical parametric oscillator (DOPO)\,\cite{drummond1980non, drummond1981} based CIMs\,\cite{wang2013coherent}, and (c) opto-electronic oscillator (OEO) based CIM\,\cite{bohm2019poor, prabhakar2023optimization}. Moreover, the DOPO-based CIM method can be further sub-classified into two architectures, depending upon their ways of obtaining the effective spin-spin interaction between the realized Ising spins: $\sigma_{i\,\left(j\right)}\in\{\pm1\}$, as specified by the coupling matrix $\J$ and the linear (Zeeman) term $\h$ representing the inhomogeneous magnetic field strength in the Ising model Hamiltonian\,\cite{kiesewetter2022coherent}: 
	\begin{align}
	    H = \frac{1}{2}\sum^{n}_{i,j}\J_{ij}\sigma_i \sigma_j + \sum^{n}_{i} \h_i \sigma_i,
	    \label{eq:IsingHamiltonian}
	\end{align}
	where $i$ (or $j$) represents each spin index or each vertex of an (un)directed spin graph. Operationally, Eq. \eqref{eq:IsingHamiltonian} denotes the cost (energy, or objective value) of a Quadratic Unconstrained Binary Optimisation (QUBO) problem\,\cite{prabhakar2023optimization}, and the $\sigma_i$s represent the binary decision variables. Among these two architectures, first is the optical delay line (ODL) architecture where a part of the time-delayed signal is redirected back into the fibre loop\,\cite{marandi2014network, takata201616}; while the other is the measurement-feedback (MFB) architecture where the feedback signal is computed electronically based on the measurement performed by a homodyne detector\,\cite{honjo2021100, mcmahon2016fully, inagaki2016coherent}. Recently, even quite a few quantum models for CIMs, based on DOPO and MFB architectures, have been theoretically developed and evaluated\,\cite{shoji2017quantum, yamamura2017quantum, yamamoto2017coherent, kiesewetter2022coherent}. Although more popular, the nonlinear DOPO-based coherent optical state generation process is neither quite scalable nor very resource efficient as it requires high power laser systems and temperature-regulated nonlinear elements, resulting into large setup footprints. On the contrary, the OEO-based CIM architecture involving self-feedback, which was first proposed in\,\cite{bohm2019poor}, offers a higher stability at a much lower form-factor. Due to its cost-effectiveness and complex nonlinear dynamics, this architecture has been lately gaining a lot of attention\,\cite{umasankar2021benchmarking, prabhakar2023optimization}. Given that these advantages are instrumental for various applications including cryptography, microwave generation, and optical neuronal computing\,\cite{bohm2019poor}, henceforth, for our rigorous convergence analyses we will focus upon this OEO-based CIM architecture.
	
	\section{OEO-based Coherent Ising Machines}\label{sec:ECIM}
	
\begin{figure}
		\centering
		\includegraphics[width=0.43\textwidth]{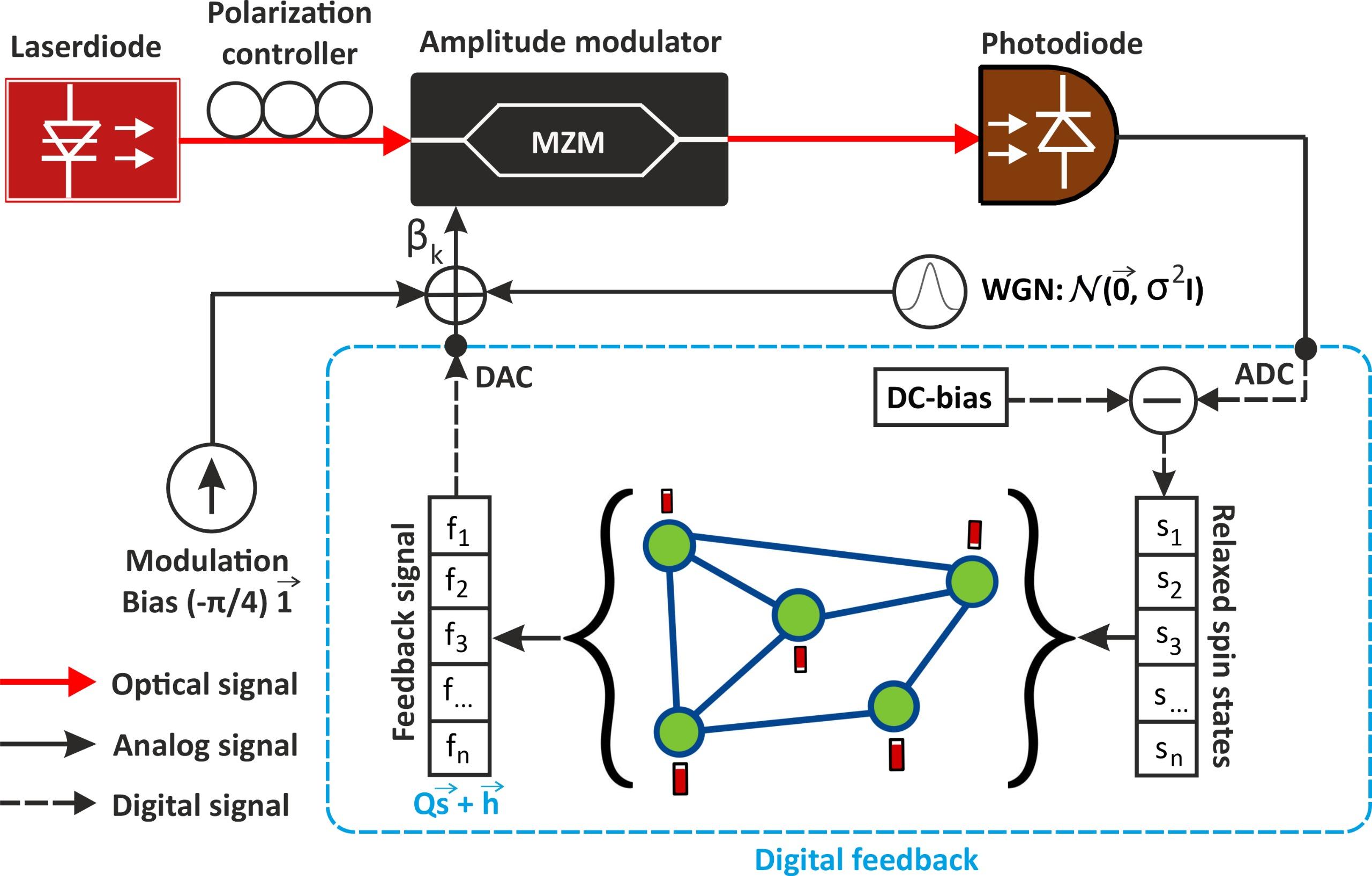}
		\caption{A conceptual schematic of our OEO-based CIM setup (adapted from \cite{bohm2019poor}). This diagram includes the modifications proposed in Sec. \ref{sec:mod} such as the use of step-sizes $\beta_k$ that are dependant on the iteration-number $k$, and the modulation of the Gaussian white noise with the step-size.}
		\label{fig:schematic}
		\vspace{-0.5cm}
	\end{figure}

	An OEO-based CIM architecture fundamentally relies upon the idea that large and regulated artificial spin networks can be created by exploiting the rich bifurcation structure of OEOs. Building upon this central idea, in this work, we have adapted the compact CIM setup, that was first demonstrated by Bohm et al. in 2019\,\cite{bohm2019poor}, as per our requirement for the convergence analysis. A conceptual schematic of this setup is presented in Fig.~\ref{fig:schematic}. As shown in the figure, a continuous wave laser beam is subjected to nonlinear opto-electronic feedback system via amplitude modulation followed by a direct photodetection of the optical signal. This enables the mapping of OEO network to a network of Ising spins. More specifically, the amplitude modulation of the input laser beam is internally performed by a Mach-Zehnder (electro-optic amplitude) modulator (MZM), set to operate in the linear regime by imparting a constant bias of $\sfrac{\pi}{4}$\,\cite{maldonado1995electro}. The polarization controller contains a combination of two quarter- and one half-wave plates, and thus converts an arbitrarily polarized beam into the desired input polarization for the amplitude modulator. In a nutshell, this CIM setup generates the artificial spins in a feedback induced pitchfork bifurcation and encodes them in the intensity of coherent states. Due to its cost-effectiveness, the OEO-based CIM was named ``a poor man's CIM" by its first demonstrators in\,\cite{bohm2019poor}. In this article, we have referred to its customized design as Economical Coherent Ising Machines and henceforth abbreviated it as ECIM. 
	In the ECIM, each binary variable $\sigma_i$ is relaxed\,\cite{linear_relaxation} to $s_i$ such that the latter takes continuous values in $[-\frac{1}{2}, \frac{1}{2}]$.The objective function, under this relaxation, takes the form:
	\begin{equation}
	    H(\s) = \frac{1}{2}\s^\top \J \s,
	\end{equation}
	which is written in the vector-form. Accordingly, the relaxed optimisation problem is given by: $\min_{\s \in [-\nicefrac{1}{2}, \nicefrac{1}{2}]^n}H(\s)$.
	Once the optimal relaxed decision variables $\s^*$ have been obtained, they are ``rounded off" to the discrete domain using an element-wise sign function: $\boldsymbol{\sigma}^* = \text{sgn}(\s^*)$.
	
	As mentioned in\,\cite{bohm2019poor}, the iterative equations of the ECIM, again in vector formats, are as follows:
	\begin{equation}\label{eq:fb}
	\fk = \alpha \sk - \beta \J\sk
	\end{equation}
	\begin{equation}\label{eq:ECIM}
	    \skp = \cos^2\left(\fk - \frac{\pi}{4}\boldsymbol{1} +\nk \right) - \frac{1}{2}\boldsymbol{1},
	\end{equation}
	where, $k$ denotes the iteration-number, $\fk$ is referred to as the feedback vector, and $\nk$ is a white Gaussian noise whose importance is discussed in detail in Sec.\,\ref{sec:noise}. It is noteworthy that the non-linear, trigonometric transfer function used here ensures that the decision variables never leave the feasible region as the iterations progress. Finally, we briefly allude to the fact that the ECIM caters only to problems where the coupling matrix $\J$ is symmetric, and the linear term $\h^\top \s$ is absent. This has been emphasised and remedied in Sec.\,\ref{sec:conv}, where unlike\,\cite{linterms}, the linear term has been included without banking on ancillary spins. Fig.\,\ref{fig:schematic} portrays a schematic of the ECIM along with the modifications proposed in this paper, such as using a $\beta_k$ which varies with iterations and modulating the noise with $\beta_k$. These modifications are expounded on in sections\,\ref{sec:conv} and\,\ref{sec:mod}.
	
	\section{Main Contributions}\label{sec:contri}
	Before proceeding further, we describe the key contributions made in this paper. The principal argument here is that, contrary to prior belief (as stated in\,\cite{McMahon}), the ECIM is not a heuristic, but a system whose performance on minimizing the relaxed objective function can be rigorously analysed. By drawing similarities between the former and noisy gradient-descent, we provide an alternate perspective to the ECIM. Further, we immaculately extend it by considering asymmetric coupling matrices and including external magnetic fields without relying on ancillary spins\,\cite{linterms}. While theoretically analysing the ECIM's performance, we mathematically justify some hithertofore empirical observations such as the effect of the coupling term $\beta$ and the role of noise. We also highlight that through the use of the ECIM in its present form, the expected value of the relaxed objective function converges in a region around the optimal value, find the dependence of the difference on the parameters, and calculate an upper bound on the number of iterations required to obtain a certain level of proximity to the optimal objective value. Finally, we address this issue by resorting to gradually-decreasing values of $\beta$ which ensures convergence to the optimal value \textit{with probability $1$}, and also find the rate of convergence of the relaxed objective function.
	
	\section{Preliminary Remarks and Assumptions}\label{sec:pre}
	In our analysis of the ECIM's convergence, we assume that $J$ has real eigenvalues, and foist the following conditions on the hyperparameters:
	\begin{enumerate}
		\item \label{alpha} In Eq. \eqref{eq:fb}, $\beta$ is restricted to be a positive quantity, while the value of $\alpha$ is fixed to $1$,
		\item The noise $\boldsymbol\zeta^{(k)}$ is sampled from the normal distribution $\mathcal{N}(\boldsymbol{0}, \sigma^2\boldsymbol{I})$, with $\boldsymbol{I}$ being the $n\times n$ identity matrix.
	\end{enumerate}
	Furthermore, the following preliminary remarks and notations may help with the ensuing analysis, which will be followed throughout the rest of the paper:
	\begin{enumerate}
		\item $\lvert\lvert \boldsymbol{v} \rvert\rvert$ denotes the $l_2$ norm of the vector $\boldsymbol{v}$,
		\item The number of relaxed decision variables is given by $n$, such that $\s \in [-\nicefrac{1}{2}, \nicefrac{1}{2}]^n$,
		\item Linear terms will henceforth be included into the objective function:
		\begin{equation}
		    E(\s) = \frac{1}{2}\s^\top \J \s + \h^\top \s,
		\end{equation}
		and the optimisation problem accordingly becomes:
		\begin{equation}
		    \min_{\s \in [-\nicefrac{1}{2}, \nicefrac{1}{2}]^n}E(\s)
		\end{equation}
		\item We use the shorthand $E^* = \min_{\s \in [-\nicefrac{1}{2}, \nicefrac{1}{2}]^n}E(\s)$,
		\item $\nabla E(\boldsymbol{s}) := \nabla_{\s} E(\boldsymbol{s}) = \frac{1}{2}(\J + \J^\top)\s+\boldsymbol{h} =: \boldsymbol{Q}\s + \h$ is used to represent the gradient of the objective function with respect to $\boldsymbol{s}$. We use $c^2$ to denote the upper-bound on the squared $l_2$ norm of the gradient for some $c < \infty$: 
		\begin{equation}\label{eq:grad_bound}
		\lvert\lvert \ges \rvert\rvert^2 \leq c^2
		\end{equation}
		Alternatively, it may be said that the function is Lipschitz-continuous.
		\item It is easy to observe that the Hessian of $E(\boldsymbol{s})$ is $\boldsymbol{Q}$. Let $\lambda_i$, $i \in [n]$, denote the eigenvalues of $\boldsymbol{J}$, with $\bar{\lambda}:=\max_i \lambda_i$. This implies that $E(\boldsymbol{s})$ is Lipschitz-smooth with a constant $\bar{\lambda}$,
		\item We define $\F_k = \{\s^{(0)}, \boldsymbol{\zeta}^{(0)}, \boldsymbol{\zeta}^{(1)}, \dots, \boldsymbol{\zeta}^{(k-1)}\}$ to be the increasing sequence of $\sigma$-fields that captures the history of the ECIM iteration process,
		\item $\mathbb{E}[X\vert \mathcal{F}]$ denotes the expectation of the random variable $X$, conditioned on the $\sigma$-field $\mathcal{F}$; $\mathbb{E}[X\vert \mathcal{F}] = X$ if $X$ is $\mathcal{F}$-measurable, and $\mathbb{E}[X\vert \mathcal{F}] = \mathbb{E}[X]$ otherwise. Additionally, if the random variable $Y$ is $\mathcal{F}$-measurable and $X$ is not, then $\mathbb{E}[XY\vert \mathcal{F}] = Y\mathbb{E}[X]$.
	\end{enumerate}
	Finally, we assume that the objective function obeys the PL inequality\,\cite{schmidt} with some $\mu > 0$, i.e.,
	\begin{equation}\label{eq:pl}
	\frac{1}{2}\lvert \lvert \nabla E(\boldsymbol{s}) \rvert \rvert ^2 \geq \mu (E(\boldsymbol{s}) - E^*),
	\end{equation}
	This assumption imposes invexity on the objective function, which is a less stringent condition than convexity\,\cite{schmidt}. It is noteworthy that equations \eqref{eq:grad_bound} and \eqref{eq:pl} do not contradict each other, unlike the argument in\,\cite{lehigh}, since in our case:
	\begin{equation}
	\lvert\lvert \s_1-\s_2 \rvert\rvert^2 \leq n, \; \forall \s_1, \s_2 \in \left[-\nicefrac{1}{2}, \nicefrac{1}{2}\right]^n
	\end{equation}

	\section{Coherent Ising Machine and Gradient-Descent}\label{sec:grad}
	With the details mentioned in Section\,\ref{sec:pre}, we begin by establishing an equivalence between Eq. \eqref{eq:ECIM} and first-order, noisy gradient-descent\,\cite{bertsekas}, the latter of which is given by\footnote{\label{fn:actual_noisy_grad}strictly speaking, the update-equation for gradient methods with errors is $\boldsymbol{s}^{(k+1)} = \boldsymbol{s}^{(k)} - \gamma_k (\nabla E(\boldsymbol{s}^{(k)}) + \boldsymbol{\zeta}^{(k)})$. We will eventually modify the ECIM's equations within this section such that they correspond exactly with the above equation.}:
	\begin{equation}\label{eq:ngd}
	\boldsymbol{s}^{(k+1)} = \boldsymbol{s}^{(k)} - \gamma_k \nabla E(\boldsymbol{s}^{(k)}) + \boldsymbol{\zeta}^{(k)},
	\end{equation}
	where, $\gamma_k>0$ is the step-size at the $k^{\text{th}}$ iteration. 
	
	On the other hand, we expand Eq. \eqref{eq:ECIM} about $\frac{\pi}{4}\boldsymbol{1}$ using Taylor's series up to the first order:
	\begin{equation}\label{eq:uwh}
	\begin{aligned}
	\skp &\approx \cos^2\left(\frac{\pi}{4}\boldsymbol{1}\right) - \left(\nabla_{\boldsymbol{\theta}} \cos^2 \boldsymbol{\theta}\right)_{\boldsymbol{\theta} = \frac{\pi}{4}\boldsymbol{1}}(\fk + \nk) - \frac{1}{2}\boldsymbol{1} \\
	&= \fk + \nk = \alpha \sk - \beta \J \sk + \nk
	\end{aligned}
	\end{equation}
	where $o(\lvert\lvert \fk + \nk \rvert\rvert)$ terms have been ignored.
	It is immediately apparent that equations \eqref{eq:ngd} and \eqref{eq:uwh} are equivalent when $\J = \J^\top$, $\h = \boldsymbol{0}$, $\alpha=1$, and $\gamma_k = \beta$. Thus, the Ising machine may be interpreted as noisy gradient-descent with $\beta$ as the step-size. Further, keeping this perspective in mind, it is natural to replace the symmetric couple matrix with an asymmetric one, and to include the external magnetic field term in the ECIM, resulting in Eq. \eqref{eq:fb} to be modified to:
	\begin{equation}\label{eq:update}
	\fk = \alpha \sk - \beta \gesk  = \alpha \sk - \beta (\boldsymbol{Q} \sk+\h ),
	\end{equation}
	which is the same equation as arrived in\,\cite{prabhakar2023optimization}, for a symmetric coupling matrix, through the use of an ancillary spin\,\cite{linterms}. Establishing a relationship between noisy gradient-descent and the ECIM leaves the latter susceptible to analysis using stochastic approximation methods, which we begin in Sec.\,\ref{sec:conv}.
	
	\section{Necessity of Noise}\label{sec:noise}
	Before delving into the convergence-analysis of the ECIM, we take a short detour to discuss the importance of noise and the role it plays in different problem-scenarios. It has been argued in\,\cite{bohm2019poor} that noise helps the ECIM escape unstable fixed points of the iteration scheme in Eq. \eqref{eq:ECIM}, which in the absence of an external magnetic field $\h$, is given by $\s = \boldsymbol{0}$. In the presence of linear terms and absence of noise, the fixed points are such that $\skp = \sk$. From Eq. \eqref{eq:update} we see that these correspond to the stationary points of $\es$ where $\ges = \boldsymbol{0}$, which in the interior of the feasible region may include global/local minima, maxima, and saddle points. The exact nature and number of such stationary points depends on the matrix $\Q$. For various types of $\Q$, we briefly discuss on the type and number of stationary points and the significance of noise in each case. It must, nonetheless, be noted that these stationary points may fall within, without, or even on the boundary of the feasible region.
	\begin{enumerate}
	    \item First, we consider $\Q$ to be a positive definite matrix. In this case, the relaxed objective function is strictly convex and admits a single stationary point which is the global minimum. If $\Q$ is positive semidefinite, then there exist multiple global minima. The presence of noise appears to be inconsequential for such convex problems.
	    \item Conversely, if $\Q$ is negative (semi)definite, then the relaxed objective function has (multiple) maxima, and the solution is known to lie on the boundary of the feasible region. In this case, noise helps the ECIM escape from the maxima in these concave problems.
	    \item Finally, if $\Q$ is has a mix of both positive and negative eigenvalues, then the problem has only saddle points, which the noise helps escape from. Here, too, the solution lies on the boundary.
	\end{enumerate}
	
	\section{Convergence Analysis}\label{sec:conv}
	We embark on the analysis of the economic CIM by first setting $\alpha=1$ and employing Taylor's series again, but this time expanding the objective function $\eskp$ about $\sk$ instead, to get:
	\begin{align*}
	\eskp &= E(\sk - \beta \gesk + \nk) \\
	&= \esk - (\beta \gesk - \nk)^\top \gesk + \\ &\phantom{==}(\beta \gesk - \nk)^\top \Q (\beta \gesk - \nk) \\
	&\leq \esk - (\beta \gesk - \nk)^\top \gesk + \\ &\phantom{==} \bar{\lambda} \lvert\lvert \beta \gesk - \nk \rvert\rvert^2 \addtocounter{equation}{1}\tag{\theequation} 
	\end{align*}
	where the inequality above is a consequence of the fact that $\boldsymbol{v}^\top \J \boldsymbol{v} \leq \bar{\lambda} \lvert\lvert \boldsymbol{v} \rvert\rvert^2$ for any $\boldsymbol{v} \in \mathbb{R}^n$. Next, we subtract $E^*$ from both sides and take expectation conditioned on $\mathcal{F}_k$, and note the following facts:
	\begin{enumerate}
		\item $E^*$, $\esk$ and $\gesk$ are $\mathcal{F}_k$-measurable, while $\nk$ is not,
		\item Further, $\mathbb{E}[\nk \vert \mathcal{F}_k]=\mathbb{E}[\nk]=\boldsymbol{0}$, $\mathbb{E}[\lvert\lvert \nk \rvert\rvert^2 \vert \mathcal{F}_k]=\mathbb{E}[\lvert\lvert \nk \rvert\rvert^2]=n\sigma^2$,
	\end{enumerate}
	applying which, we arrive at: 
	\begin{multline}
	\!\!\!\!\cedeskp \leq [\desk] - \beta\lvert\lvert \gesk \rvert\rvert^2 \\ + \bar{\lambda}(\beta^2 \lvert\lvert \gesk \rvert\rvert^2 + n\sigma^2)
	\end{multline}
	which, on using equations \eqref{eq:grad_bound} and \eqref{eq:pl}, becomes:
	\begin{multline}\label{eq:cond_exp}
	\cedeskp \leq (1-2\beta\mu)[\desk] \\ + \bar{\lambda}(\beta^2 c^2 + n\sigma^2)
	\end{multline}
	Finally, we take unconditional expectation on both sides of the equation above to obtain:
	\begin{equation}\label{eq:uncond_exp}
	\edeskp \leq (1-2\beta\mu)\edesk + \bar{\lambda}(\beta^2 c^2 + n\sigma^2)
	\end{equation}
	At this point, we highlight a minor problem with the convergence of the ECIM and hypothesise that $\ex[\esk]$ cannot get arbitrarily close to $E^*$ with a suitably-chosen, constant step-size $\beta$.
	
	We formalise our claim as:
	\begin{equation}\label{eq:opt_gap1}
	\liminf_{k \rightarrow \infty}\edesk \leq \frac{\bar{\lambda}}{2\mu}\left(\beta c^2 + \frac{n\sigma^2}{\beta}\right)
	\end{equation}
	and set about proving it by contradiction, along the lines of that in\,\cite{nedic1}. To do so, we assume the above statement to be false, i.e., for some $\epsilon >0$:
	\begin{equation}\label{eq:add1}
	\liminf_{k \rightarrow \infty}\edesk \geq \frac{\bar{\lambda}}{2\mu}\left(\beta c^2 + \frac{n\sigma^2}{\beta}\right) + 2\epsilon
	\end{equation}
	Now suppose there exists a $k_0$ which is large enough such that for all $k \geq k_0$, we have:
	\begin{equation}\label{eq:add2}
	\ex[\esk] \geq \liminf_{k \rightarrow \infty} \ex[\esk] - \epsilon
	\end{equation}
	Adding equations \eqref{eq:add1} and \eqref{eq:add2}:
	\begin{equation}
	\edesk \geq \frac{\bar{\lambda}}{2\mu}\left(\beta c^2 + \frac{n\sigma^2}{\beta}\right) + \epsilon
	\end{equation}
	Substituting this back into Eq. \eqref{eq:uncond_exp}, we end up with the following recurrence inequality:
	\begin{equation}
	\edeskp \leq \edesk - 2\beta\mu\epsilon
	\end{equation}
	We then unroll the above inequality for $K$ steps to get:
	\begin{equation}\label{eq:unroll}
	\edesk \leq \ex[E(s^{(k-K)})-E^*] - 2\beta\mu\epsilon K
	\end{equation}
	But at this recall we note that $E^*$ is the optimal objective value, and hence, $(\desk) \geq 0$ for any $k$. Further, the objective values are bounded, which means that the above recurrence cannot continue indefinitely. This leads to the desired contradiction, thus completing the proof of our claim.
	
	A keen observation of Eq. \eqref{eq:opt_gap1} confirms our initial speculation that one may not be able to select a suitable low $\beta$ to get as close to the optimal objective value as required. If a low value of $\beta$ is chosen, then the second term in the equation increases in magnitude, according us no benefit towards reducing the expected gap between $\esk$ and $E^*$. This second term has an even greater effect and significance in larger problems with a larger number of decision variables $n$, indicating a poor scaling of the ECIM in its present form. This helps explain the observation in Fig. 3(c) of\,\cite{bohm2021} that the TTS (``Time-to-Solution", as defined in the same reference) first decreases with increasing $\beta$ and then, counter-intuitively, starts increasing; and that there is a minute increase in TTS as $\sigma^2$ is increased. Granted, that one may opt to reduce the noise-variance to counter the effect of $\nicefrac{n}{\beta}$, but this leaves the ECIM prone to get stuck at local minima, saddle points, or maxima, as discussed in Sec.\,\ref{sec:noise}.
	
	\section{Proposed Modifications}\label{sec:mod}
	To offset the difficulties in convergence of the ECIM, as discussed at the end of the last section, we first present a slight shift in perspective of the ECIM's update equations, which will offer no immediate benefit, but facilitate the modifications that will be communicated in the latter half of this section. The proposition is simple: that the white Gaussian noise be scaled with the step-size $\beta$ such that equations \eqref{eq:fb} and \eqref{eq:ECIM} are modified to:
	\begin{equation}
	\fk = \alpha \sk - \beta_k\left(\gesk - \nk\right),
	\end{equation}
	and
	\begin{equation}
	\skp = \cos^2\left(\fk -\frac{\pi}{4}\boldsymbol{1}\right) - \frac{1}{2}\boldsymbol{1},
	\end{equation}
	respectively, where $\beta_k$ is the step-size at iteration $k$. At face-value, nothing appears to have changed, but we posit that with gradually-decreasing step-sizes, the ECIM converges to the optimal objective value ``almost surely".
	
	However, before discussing about diminishing step-sizes, we repeat the analysis in Sec.\,\ref{sec:conv} with $\beta=\beta_k$ to arrive at the modified versions of equations \eqref{eq:cond_exp} and \eqref{eq:uncond_exp1}:
	\begin{multline}\label{eq:cond_exp1}
	\cedeskp \leq (1-2\beta_k\mu)[\desk] \\ + \bar{\lambda}\beta_k^2(c^2 + n\sigma^2),
	\end{multline}
	and
	\begin{multline}\label{eq:uncond_exp1}
	\edeskp \leq (1-2\beta_k\mu)\edesk \\ + \bar{\lambda}\beta_k^2(c^2 + n\sigma^2).
	\end{multline}
	
	\subsection{With a Constant Step-Size}
	With the modified equations presented above, using techniques similar to the ones employed previously, it may be shown that with constant step-sizes $\beta_k=\beta$:
	\begin{equation}
	\liminf_{k \rightarrow \infty}\edesk \leq \frac{\bar{\lambda}}{2\mu}\beta(c^2 + n\sigma^2)
	\end{equation}
	It is important to note here that the step-size $\beta$ may be controlled by the user to make $\esk$ get arbitrarily close to $E^*$. Further, it is also possible to find an estimate of the number of iterations $\kappa$ required such that:
	\begin{equation}
	\min_{0\leq k\leq \kappa} \edesk \leq \frac{\bar{\lambda}}{2\mu}\beta(c^2 + n\sigma^2) + \epsilon,
	\end{equation}
	where $\epsilon > 0$ and
	\begin{equation}
	\kappa \leq \bigg\lfloor \frac{E(\boldsymbol{s}^{(0)})-E^*}{2\beta\mu\epsilon} \bigg\rfloor
	\end{equation}
	This is easily obtained by setting both $k$ and $K$ to $\kappa$ in Eq. \eqref{eq:unroll} such that:
	\begin{equation}
	0 \leq \edesk \leq [E(\boldsymbol{s}^{(0)})-E^*] - 2\beta\mu\epsilon\kappa
	\end{equation}
	We would like to reiterate that merely scaling the noise term with the step-size is equivalent to using noise with a lower variance and does not provide any additional benefits with constant step-sizes, but the variation in perspective allows us to inspect the ECIM with decreasing step-sizes, that we now foray into.
	
	\subsection{With Decreasing Step-Sizes}
	Here, we prove that it is possible to drive $\esk$ to $E^*$ with probability $1$ with the help of constantly decreasing step-sizes. As is customary in machine learning literature, we consider step-sizes of the following form:
	\begin{equation}\label{eq:beta}
	\sum_{k=0}^{\infty} \beta_k = \infty \;\; \text{   and   } \;\; \sum_{k=0}^{\infty} \beta_k^2 < \infty
	\end{equation}
	One popular example of a sequence of step-sizes that follows the above criteria is $\beta_k = \nicefrac{\beta_0}{(k+1)^r}$ where, $\beta_0>0$ and $r \in (0.5,1]$.
	
	With these step-sizes, we note that in Eq. \eqref{eq:cond_exp1}, the terms $(\edesk)$, $2\beta_k\mu(\edesk)$ and $\bar{\lambda}\beta_k^2(c^2+n\sigma^2)$ are non-negative and $\mathcal{F}_k$-measurable. Further, the latter sequence is summable, i.e.:
	\begin{equation}
	\bar{\lambda}(c^2+n\sigma^2)\sum_{k=0}^{\infty} \beta_k^2 < \infty.
	\end{equation}
	This enables us to invoke the Supermartingale Convergence Theorem (SMCT, as mentioned in Proposition 2.2 of\,\cite{smct}), which guarantees that:
	\begin{enumerate}
		\item $\desk$ converges to some non-negative value almost surely, and
		\item \begin{equation}\label{eq:smct_sum}
		\sum_{k=0}^{\infty}2\beta_k\mu (\desk) < \infty \text{ w.p. 1}
		\end{equation}
	\end{enumerate}
	
	Now we show that not only does $(\desk)$ converge to a non-negative value, but that it converges to $0$ almost surely, essentially meaning that $\esk \rightarrow E^*$ a.s. To demonstrate this, we fall back on proof by contradiction again, and consider an $\epsilon>0$ and a $k_0>0$ such that for all $k\geq k_0$, 
	\begin{equation}\label{eq:assum}
	(\desk)\geq \epsilon, 
	\end{equation}
	which leads to:
	\begin{equation}
	\sum_{k=k_0}^{\infty}2\beta_k\mu (\desk) \geq 2\mu\epsilon\sum_{k=k_0}^{\infty}\beta_k,
	\end{equation}
	the right hand side of which is $\infty$ as per the first part of Eq. \eqref{eq:beta}. But this contradicts the consequence of SMCT shown in Eq. \eqref{eq:smct_sum}, which is known to be true. Thus, the assumption in Eq. \eqref{eq:assum} is false and $(\desk)\rightarrow 0$ almost surely, thus concluding our proof.
	
	Further, it is also possible to get the rate of decrease of $\edesk$ with diminishing step-sizes by a direct application of Lemma 2.1 in\,\cite{nedic2} on Eq. \eqref{eq:uncond_exp1}.
	
	\section{Conclusion and Outlook}\label{sec:conclusion}
	In this paper, we demonstrated that ECIMs are not heuristics, unlike they were previously believed to be. This was, however, done in a restricted setting where only a linear approximation of a single trigonometric transfer function was considered, and $\alpha$ was fixed to $1$. We conjecture that the proofs presented here may be extended to the more holistic case that generalises these aforementioned constraints. Besides, the performance of the ECIM was theoretically analysed only for the instance where the binary decision variables were relaxed to take continuous values. As future work, the analysis may be carried out by considering a conjunction of Ising Machines and rounding-off methods (such as branch-and-bound\,\cite{bnb}), the latter of which may be used to convert the decision variables back to the discrete domain. Overall, this paper paves the way for rigorous analysis of Ising Machines which may prove to be compelling alternatives towards solving QUBO problems.
	
	\section{Acknowledgement}\label{sec:ack}
	We would like to thank Mr. Anil Sharma, Mr. Vidyut Navelkar, Mr. C. V. Sridhar, Mr. Godfrey Mathais, and Dr. Anirban Mukherjee from the Corporate Incubation team at Tata Consultancy Services, and Dr. M. Girish Chandra from TCS Research for their unwavering support throughout the course of this work. SP and SC also acknowledge the initial discussions with Prof. Anil Prabhakar from IIT Madras.
	
	\balance
	\bibliographystyle{IEEEtran}
    \bibliography{refs}

	
\end{document}